\documentclass[a4paper,11pt]{article}
\usepackage{comment}
\usepackage{pos}

\newtheorem*{conjecture*}{Conjecture}

\renewcommand{\Re}{\mathrm{Re}\;}
\renewcommand{\Im}{\mathrm{Im}\;}

\title{Normalizing flows for the real-time sign problem}
\author*[a]{Yukari Yamauchi}
\author[b]{Scott Lawrence}
\affiliation[a]{Department of Physics, University of Maryland,\\ College Park, Maryland 20742, USA}
\affiliation[b]{Department of Physics, University of Colorado,\\ Boulder, Colorado 80309, USA}

\emailAdd{yyukari@terpmail.umd.edu}
\abstract{We discuss the application of normalizing flows to bosonic lattice field theories with real-time sign problems. A normalizing flow, once it is found for such a lattice field theory, is guaranteed to solve its sign problem. We argue for the existence of normalizing flows for bosonic lattice field theories in the Schwinger-Keldish formalism in a few ways. We then discuss how this existence is a specific feature of bosonic theories: such arguments break down for fermionic systems, whether at finite density or in real-time.}

\FullConference{%
 The 38th International Symposium on Lattice Field Theory, LATTICE2021
  26th-30th July, 2021
  Zoom/Gather@Massachusetts Institute of Technology
}

\begin{document}
\maketitle

\section{Introduction}
Lattice calculations have successfully revealed many aspects of quantum chromodynamics (QCD) by performing non-perturbative calculations of QCD observables in the path integral formalism from first principles. However, there are still many kinds of observables yet to be addressed by lattice QCD due to the so-called sign problem. One setting where a sign problem occurs is in the calculations of lattice field theories done in Minkowski spacetime. The path integral
\begin{equation}
    Z =\int \mathcal{D}[\phi] e^{-S(\phi)}
\end{equation}
for Minkowski spacetime is defined with a complex-valued action $S$. Here $\phi$ represents all lattice degrees of freedom, and we assume for simplicity that they take values in $\mathbb{R}^N$. As the ``Boltzmann factor'' $e^{-S}$ is also complex-valued, one cannot simply regard it as a probability distribution of configurations $\phi$, which is the standard interpretation needed to apply Markov chain Monte Carlo (MCMC) sampling. One way to get around the problem is to define the so-called quenched distribution function $e^{-\Re S}$. With this quenched distribution, the expectation value of observables $\mathcal{O}(\phi)$ on a lattice can be computed as
\begin{equation}\label{eq:qp}
    \langle \mathcal{O(\phi)} \rangle = \frac{\int \mathcal{D}[\phi]\; e^{-S}\; \mathcal{O(\phi)} / \int \mathcal{D}[\phi]\; e^{-\Re S}}{\int \mathcal{D}\;[\phi] e^{-S}/\int \mathcal{D}[\phi] \;e^{-\Re S}} = \frac{\langle \mathcal{O}(\phi) \;e^{-i \Im S} \rangle_Q}{\langle e^{-i \Im S} \rangle_Q}
    \text,
\end{equation}
where $\langle\cdot\rangle_Q$ denotes that the expectation values are evaluated with the quenched distribution. Both numerator and denominator in (\ref{eq:qp}) are numerically challenging to compute; individually, each suffers from a severe signal-to-noise problem. Consider for example the denominator, termed the average sign and often denoted as $\langle\sigma\rangle$. The average sign is (in the presence of a sign problem) strictly less than $1$, and scales exponentially with the volume: $\langle\sigma\rangle \sim e^{-V}$. However, each sample is a complex number of magnitude $1$, and therefore the MCMC sampling needs $\sim e^{2V}$ samples to resolve the average sign from $0$. Such a lattice calculation is not scalable for systems of physical interest. This is the sign problem. A similar issue occurs for many fermionic models, including QCD at non-zero baryon density.

While there are several proposal for alleviating sign problems proposed, one long-standing method which successfully tamed sign problems for some models of our interest is the so-called manifold deformation method~\cite{ref:md}. The idea of the method is simple: we deform the contour of integration is the path integral, $\mathbb{R}^N$ to the complex plane $\mathbb{C}^N$ of the field variables $\phi$, aiming for a larger average sign, and thus milder sign problem. When the average sign is exactly $1$, the sign problem is \emph{solved} --- the average sign stays exactly $1$ no matter how large the size of the lattice becomes. 

The equality of expectation values before and after a contour deformation is guaranteed~\cite{ref:thesis} by Cauchy's integral theorem. To be precise, let $\mathcal M$ be a contour obtainable by a continuous deformation of $\mathbb R^N$, where the deformation passes only through regions in which both $e^{-S(\phi)}$ and $\mathcal O(\phi) e^{-S(\phi)}$ are holomorphic functions of $\phi$. As long as the asymptotic behavior of the contour does not change, the expectation value $\langle \mathcal O\rangle$ evaluated on $\mathcal M$ will equal the physical expectation value.

% As an example of manifold deformation, for the following ``one-site model'' ($\phi \in \mathbb{R}$) 
% \begin{equation}\label{phi4}
%     S = i m \phi^2 + i \lambda \phi^4
% \end{equation}
% one can numerically search for the deformed contour in the complex plane of $\phi$. In FIG.~\todo, we plotted examples of such contours with different average sign. 

The main result of our paper~\cite{ref:nf} was that we showed that such perfect manifolds (with average sign of 1) exist, at least for bosonic theories in Schwinger-Keldysh formalism. In this proceeding, we review the argument for the existence of perfect manifolds given in~\cite{ref:nf} in Section~\ref{sec2}, and then give a separate, new argument in Section~\ref{sec3}. Finally, in Section~\ref{sec4}, we show how both arguments fail for specific theories, including fermionic systems at finite density -- it remains unclear if perfect manifolds exist in those cases.% We conclude in Section~\ref{con}.

\section{Perfect contours from the holomorphic gradient flow}\label{sec2}
In this section, we review the construction of a perfect manifold described in~\cite{ref:nf}. One approach to constructing contours with improved sign problems is via the holomorphic gradient flow~\cite{ref:hgf}, in which we deform the contour by changing each point $\phi$ via the following differential equation with the flow time $t$:
\begin{equation}\label{eq:gf}
\frac{\mathrm{d} \phi}{\mathrm{d} t} = \overline{\frac{\partial S}{\partial \phi}}\text.
\end{equation}
At each flow time $t$, we obtain a new integration contour. The holomorphic gradient flow is often applied starting from the real axis. In this case, at early flow time, the holomorphic gradient flow is guaranteed to improve the sign problem, as is discussed in~\cite{ref:nf}. This can be seen by looking at how the action changes with flow time:
\begin{equation}
    \frac{\mathrm{d} S}{\mathrm{d} t} = \left|\frac{\partial S}{\partial \phi}\right|^2\text.
\end{equation}
The real part of the action increases, making the denominator of the average sign smaller. In the meantime, the numerator of the average sign does not change when the contour satisfies the three conditions discussed above. Thus the overall average sign increases. At later times, the average sign is not guaranteed to improve due to two additional contributions: the ``local" sign problem due to the Jacobian coming from the deformed contour, and the ``global" sign problem due to the zeros of the Boltzmann factor. In this section, we show how one can in principle obtain a manifold that is at least locally perfect by erasing the local sign problem.

To construct a locally perfect manifold, we make use of the fact that the holomorphic gradient flow (\ref{eq:gf}) always improves the average sign right after the manifold flows off the real plane. We flow the manifold (initially $\mathbb R^N$) for some small time $\epsilon$; this improves the average sign by a little bit. We then parametrize the new contour by the real plane $\mathbb{R}^N$. This defines a new effective action on $\mathbb{R}^N$, which we use to flow the manifold again for time $\epsilon$, further improving the sign problem. We repeat the process many time until a fixed point is obtained --- we will see below that such a fixed point must at least have no \emph{local} sign problem. In the following, we describe the process in more detail, and finally discuss the properties of the fixed-point contour.

We start with the action $S(\phi)$. As the first step, we flow the original integration manifold $\mathbb{R}^N$ with the holomorphic gradient flow, Eq.~(\ref{eq:gf}), until the flow time $t = \epsilon$. All $\phi \in \mathbb{R}^N$ will be moved to $\phi_1 \in \mathbb{C}^N$ according to
\begin{equation}
        \phi_1(\phi) = \phi + \epsilon \overline{\frac{\partial S}{\partial \phi}}
    \text.
\end{equation}
This map $\phi_1(\phi)$ defines a new integration contour $\mathcal{M}_1$. The new contour $\mathcal{M}_1$ is \emph{parametrized} by $\phi \in \mathbb{R}^N$ via $\phi_1(\phi)$. This parametrization defines the effective aciton $S_1$:
\begin{equation}
S_1(\phi) = S[\phi_1(\phi)] -
\log \det \left(
1 +
\epsilon \frac{\partial}{\partial \phi} \left.\overline{\frac{\partial S}{\partial \phi}}\right|_{\bar \phi}
\right)
\text.
\end{equation}
Here a serious complication arises. The nice properties of the holomorphic gradient flow stem chiefly from the fact that the action being used to flow is a holomorphic function; this effective action $S_1$, however, is clearly not holomorphic! The situation is remedied by realizing that the procedure above really only defined $S_1$ on $\mathbb R^N$. We can choose to defined the behavior of the effective action on $\mathbb C^N$ any compatible way we like; in particular, we choose it to be the analytic continuation of the function of the real plane.

Now that $S_1(\phi)$ is the analytic function of $\phi$ on $\mathbb{R}^N$, we can let the manifold $\mathbb{R}^N$ flow via Eq.~(\ref{eq:gf}) again. We obtain a map $\phi_2(\phi)$ which defines a new manifold, and again is guaranteed to improve the sign problem. We repeat the small flow by $\epsilon$ and the projection back to $\mathbb{R}^N$ many times, improving sign at each step. What will happen after we perform many steps? 

When the manifold reaches a fixed point $\mathcal{M}_f$, after $\mathcal{M}_f$ is projected onto $\mathbb{R}^N$, the next flow does not move the manifold off of the real plane (although individual points will flow \emph{within} $\mathbb R^N$). In other words, its effective action $S_f(\phi)$ satisfies
\begin{equation}\label{eq:noflu}
    \overline{\frac{\partial S_{f}}{\partial\phi}} \in \mathbb{R}^N,
\end{equation}
except at the singularities of $S_f$. In other words, on the perfect manifold, the effective Boltzmann factor $e^{-S_f(\phi)}$ has no phase fluctuations:
\begin{equation}
   e^{-S_f}\;\frac{\partial \Im S_f}{\partial\phi} = 0\;\;\; 
\end{equation}
So the phase $\Im S$ is guaranteed to be constant except at singularities of $S_f$ on $\mathcal{M}_f$. Any remaining sign problem must come from \emph{global} cancellations between segments of the contour, separated by zeros of the Boltzmann factor.

\begin{figure}[h]
    \centering
    \includegraphics[width=0.5\textwidth]{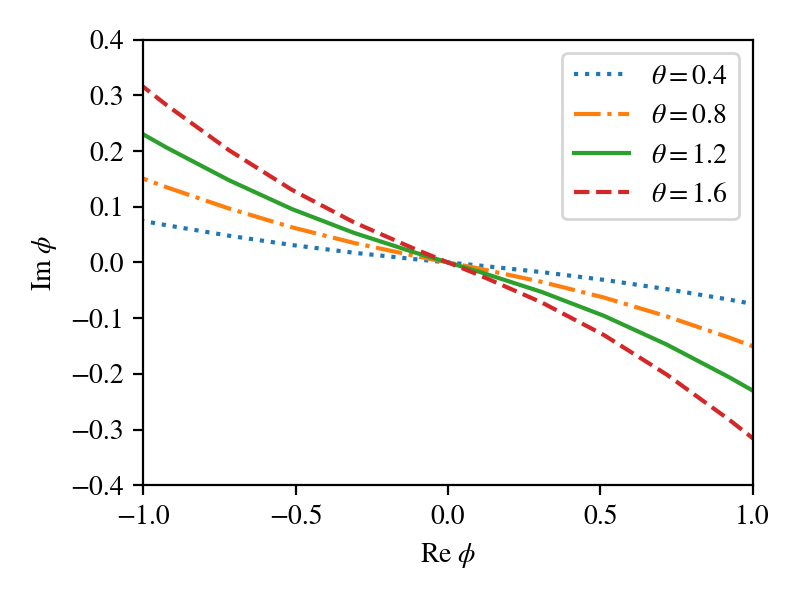}
    \caption{Integration contours with the average sign numerically measured to be $\langle \sigma\rangle > 1 - 10^{-5}$ for the model (\ref{eq:1d}) with various $\theta$.}
    \label{fig:1d}
\end{figure}

We can demonstrate the existence of perfect manifolds in the following ``one-site model'' ($\phi \in \mathbb{R}$):
\begin{equation}\label{eq:1d}
    S = \phi^2  + e^{i\theta} \phi^4\text.
\end{equation}
When $\theta=0$, the model doesn't have a sign problem. For a non-zero $\theta\in\mathbb{R}$, one can numerically search for (approximately) perfect manifolds in the complex plane of $\phi$. In Figure~\ref{fig:1d}, we plot examples of such contours with several choices of $\theta$. Note that the perfect manifold appears to change continuously as a function of $\theta$.

The argument above suggested only that \emph{locally} perfect contours exist; meanwhile, Figure~\ref{fig:1d} and similar numerical experiments in few dimensions suggest the stronger statement that globally perfect contours are available. We can strengthen the above argument to address globally perfect contours by making use of the observation that the perfect contour changes continuously as the action parameters are varied. Specifically, assume the following:
\begin{conjecture*}
For every continuous family of actions $S_\lambda$, there is a continuous family of contours $\mathcal M_\lambda$ such that $\mathcal M_\lambda$ is a locally perfect contour for $S_\lambda$. Moreover, any contour $M_0$ which is locally perfect is part of some such continuous family.
\end{conjecture*}
At first glance, this is only a minor assumption on top of the argument given above for the existence of locally perfect contours. However, by constraining the behavior of locally perfect contours as the action parameters $\lambda$ are changed, it implies the existence of \emph{globally} perfect manifolds for a broad class of actions. In particular, take $S_\lambda$ to be the Schwinger-Keldysh action for bosons with a $\phi^4$ coupling of $\lambda$. When $\lambda = 0$, a globally perfect contour is easily found, as the path integral is Gaussian. This globally perfect contour is continuously connected to locally perfect contours at larger $\lambda$. However, in the absence of zeros of the Boltzmann factor $e^{-S}$, the act of creating multiple segments of the contour with different phases is discontinuous. From this we conclude that the locally perfect contours at finite $\lambda$ must in fact be globally perfect.

\section{Complex normalizing flows}\label{sec3}

Another way to view perfect contours is as an analytic continuation of a normalizing flow. A normalizing flow is a map $\phi : \mathbb R^N \rightarrow \mathbb R^N$ obeying
\begin{equation}\label{eq:flow}
    \det \left( \frac{\partial \phi}{\partial x}  \right) e^{-S(\phi(x))} = \mathcal{N} e^{-x\cdot x/2}
    \text.
\end{equation}
The normalization constant $\mathcal N$ is given by the partition function of the physical model, and will drop out of all equations in this discussion.

When the action is not complex-valued, and has no sign problem, a normalizing flow is guaranteed to exist~\cite{ref:villani}, and for $N>1$, is far from unique. Normalizing flows have been successfully applied in lattice field theory to accelerate the MCMC sampling~\cite{ref:MSA,ref:KAN}. The case of a complex-valued action (a ``complex normalizing flow'') is discussed in~\cite{ref:nf}: in short, such a normalizing flow yields a perfect contour given by the image of $\mathbb R^N$ in $\mathbb C^N$ under the map $\phi$. Thus, for the complex normalizing flow to exist, a perfect contour has to exist. The converse also holds: when a perfect contour exists, a normalizing flow (generally far from unique) is also guaranteed to exist.

In fact, not only are perfect contours connected to normalizing flows in the abstract sense of being normalizing flows ``in the complex case'', but individual perfect contours can be obtained concretely by analytic continuation of normalizing flows, viewed as functions of the action parameters. To see how this works, let us consider \emph{perturbative} normalizing flows of the $\phi^4$ scalar theory, given by the action:
\begin{equation}
    S = \sum_{ij} \phi_i M_{ij} \phi_j + \lambda \sum_i \Lambda_i \phi_i^4\text.
\end{equation}
Here we fix $\lambda$, the magnitude of the strength of the coupling. We consider $M$ and $\Lambda$ as the action parameters, and will analytically continue a normalizing flow in the space of $M, \Lambda$ to obtain a normalizing flow for a choice of these parameters for the $\phi^4$ scalar theory in the Schwinger-Keldysh model which possesses a sign problem.

As it was detailed in~\cite{ref:nf}, we can solve the differential equation, Eq.~(\ref{eq:flow}), for the map $\phi(x)$ in either the weak- or strong-coupling limit analytically. At weak coupling, a normalizing flow is given by
\begin{equation}\label{eq:pf1-sk}
    \phi^{\mathrm{weak}}_i(x) = x_i -\lambda \sum_{j} \left[
    \frac 1 2 M^{-1}_{ij} \Lambda_j x^3_j +
    \frac 3 4 M^{-1}_{ij} M^{-1}_{jj} \Lambda_j x_j
    \right]
    \text.
\end{equation}
Thus the perturbative flow is the analytic function of the action parameter $M, \Lambda$ except at vanishing $\det M$. Perfect contours are obtained by using complex values of $M_{ij}$ and $\Lambda_i$. Unfortunately, the weak-coupling expansion is also an expansion in small $\phi$, and therefore the integration contours obtained by analytic continuation do not lie in the correct homology class.

To study the behavior of integration contours at large $\phi$, we can also find the perturbative flow in the strong coupling limit. We will write the map as the sequence of four maps:
 \begin{equation}
 \phi^{\mathrm{strong}}(x) =
 \left[F_4\circ F_3\circ F_2\circ F_1
 \right](x)
 \text.
 \end{equation}
Firstly, $F_1:\mathbb{R}\rightarrow\mathbb{R}$ maps $e^{-x_i^2/2}$ to $e^{-x_i^4}$ for each site $i$ . This map has no dependence on $M, \Lambda$.
The second map $y_i=F_2(x_i)$ then maps the distribution $e^{-x_i^4}$ to $e^{-\Lambda y_i^4}$ via the rotation and rescaling of the complex plane: 
\begin{equation}
F_2(x_i) = x_i/\Lambda_i^{1/4}\text.
\end{equation}
This map is thus analytic in $M, \Lambda$ except at vanishing $\Lambda$.

The map $\vec z = F_3(\vec y)$ then transforms the distribution $e^{- \sum_i \Lambda_iy_i^4}$ to $e^{-S'(\vec z)}$ with 
\begin{equation}\label{eq:rescaledS}
        S'(\vec z) = \sum_i\Lambda_i z_{i}^4 + \frac{1}{\sqrt{\lambda}}\sum_{ij}z_{i} M_{ij}z_{j}
        \text.
\end{equation}
correctly up to first order in $\frac{1}{\sqrt{\lambda}}$. The perturbative piece $\delta(\vec y)$ of the map $z_i = y_i + \frac{1}{\sqrt{\lambda}} \delta_i(\vec y)$ is
\begin{eqnarray}\label{eq:nf-strong}
\delta_{i}(\vec y) &=& e^{\Lambda_i y_i^4}M_{ii}\left[-\frac{ y_i^3 \Gamma[\frac{3}{4},\Lambda_i y_i^{4}]}{4(\Lambda_i y_i^{4})^{3/4}}+ \frac{\langle y_i^{2} \rangle y_i \Gamma[\frac{1}{4},\Lambda_i y_i^4]}{4(\Lambda_i y_i^{4})^{1/4}}\right]\nonumber\\   
&+& \sum_{j \in \{i-1,i+1\}}
    e^{\Lambda_i y_i^4}\frac{\sqrt{\pi}}{4\sqrt{\Lambda_i}}\left[ \mathrm{Erf}(\sqrt{\Lambda_i}y_i^{2}) - C\right] M_{ij} y_{j}
    \text.
\end{eqnarray}

In the expression of the map $\delta(\psi)$, $(\cdot)^{1/4}$ means that we take the principle fourth root. Regarding the constant $C$, a choice of $C=1$ gives a map which vanishes at $\psi_i \rightarrow\infty$ and is oscillation-free.  Finally, the map $\phi_i=F_4(z_i)$ rescales the complex plane again to obtain the desired distribution $e^{-S(\phi)}$: $F_4(z_i)= z_i/\lambda^{1/4}$. This map, like the first, has no dependence on $M$ or $\Lambda$.

Thus, we have found a normalizing flow, in our strong-coupling expansion, is an analytic function of $M, \Lambda$ (except at vanishing $\Lambda$). This flow can trivially be used at complex $M,\Lambda$, yielding perfect contours, again to first order in the expansion.

\section{Boltzmann factor zeros}\label{sec4}
When a manifold intersects with zeros of the Boltzmann factor (including those at infinity), the manifold, while locally perfect, may fail to be globally perfect. In the context of normalizing flows, at the singularities $\phi(x_0)=\phi_0$ where $e^{-S(\phi_0)}=0$, one of the following has to happen:
\begin{enumerate}
    \item The Jacobian diverges: $\det \left( \frac{\partial \phi}{\partial x}  \right)_{\phi\rightarrow\phi_0}= \infty$ while $x_0$ stays finite.
    \item The map $\phi$ will send the point $x_0$ to infinity, $|x_0| \rightarrow \infty$.
\end{enumerate}
This section is devoted to zeros of the Boltzmann factor that lie away from infinity; in other words, we will investigate the first case in detail.

\begin{figure}[h]
    \centering
    \includegraphics[width=0.47\linewidth]{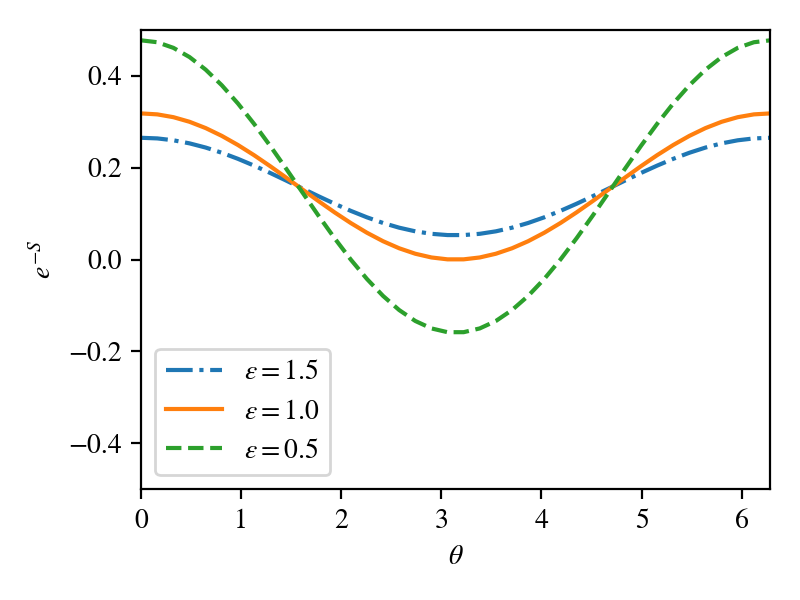}\hspace{0.04\linewidth}
    \includegraphics[width=0.47\linewidth]{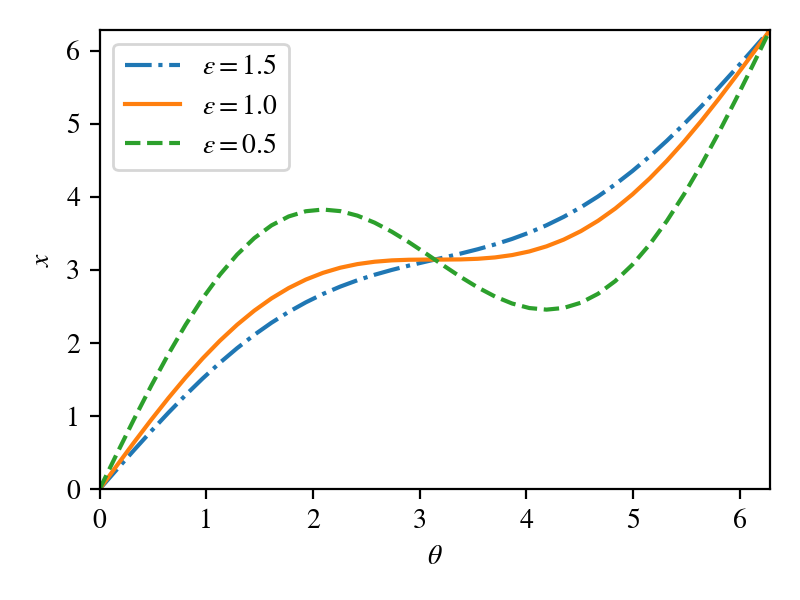}
    \caption{The left panel: The Boltzmann factor Eq.~(\ref{eq:cos}) with $\epsilon=1.5, 1.0, 0.5$. The right panel: the normalizing flow $x(\theta)$ with $\epsilon = 1.5,1.0,0.5$.}
    \label{fig:cos}
\end{figure}

A good concrete example is given by the following one-dimensional Boltzmann factor:
\begin{equation}\label{eq:cos}
    e^{-S_\epsilon(\theta)} = \cos(\theta) + \epsilon,\;\; \theta \in [0,2\pi]
\end{equation}
The uncomplexified domain of integration is the circle; the corresponding complex space is $S^1 \times \mathbb R$. The parameter of the action, $\epsilon$, can be any complex number. A sign problem is already be obtained for real $\epsilon \in (-1,1)$; for real $\epsilon$ outside of this range there is no sign problem. The left-hand panel of Figure~\ref{fig:cos} shows the Boltzmann factor for several values of $\epsilon$.

For all $\epsilon \in \mathbb R$, the real line is a locally perfect contour. When $\epsilon > 1$, the zeros do not intersect this contour, and it is therefore globally perfect. As $\epsilon$ is lowered, the zeros come down, intersect the contour for the first time at $\epsilon = 1$, and thence create global cancellations when $\epsilon < 1$ (but still positive). For all $\epsilon \in \mathbb R$, the locally perfect contour of real $\theta$ is also the best possible contour available. In other words, when $\epsilon \in (-1,1)$, there is no globally perfect contour.

Let us now consider normalizing flows for this action. As the action is defined on the finite range of $\theta$, we choose the normalizing flow to be a map from the distribution Eq.~(\ref{eq:cos}) to the uniform distribution on $x \in [0,2\pi)$; that is, $\theta(x)$ is a solution to
\begin{equation}\label{eq:nfcos}
    \frac{\mathrm{d}\theta(x)}{\mathrm{d}x}\;\frac{(\cos(\theta(x))+\epsilon)}{\epsilon} = 1\text.
\end{equation}
The solution is unique up to an arbitrary shift of $\theta$ (or equivalently of $x$). The inverse of such a map $\theta(x)$ for real $\epsilon > 1$ is
\begin{equation}
    x(\theta) = \left( \sin(\theta) + \epsilon \theta \right)/\epsilon
    \text.
\end{equation}
In the right panel of Figure~\ref{fig:cos}, the map $x(\theta)$ with $\epsilon = 1.5$ is shown. Now, we lower $\epsilon$ towards 1. We find that when $\epsilon=1$, at $\theta=\pi$ where the Boltzmann distribution vanishes, the Jacobian $\mathrm{d} \theta / \mathrm{d} x$ diverges such that Eq.~(\ref{eq:nfcos}) holds. Now, we keep lowering $\epsilon$ and get into the regime with a sign problem: here, the flow $\theta(x)$ is no longer single-valued, reflecting the fact that there is no globally perfect contour.

\begin{figure}
    \centering
    \includegraphics[width=0.55\textwidth]{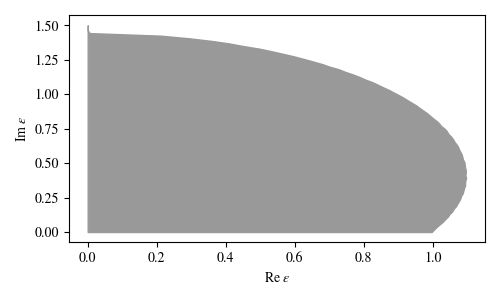}
    \caption{The region of $\epsilon$, in the first quadrant of the complex plane, for which the Boltzmann factor (\ref{eq:cos}) lacks a globally perfect contour.}
    \label{fig:perfect-region}
\end{figure}

We can also consider what happens at complex $\epsilon$. Here, both the (locally) perfect contour and the zeros move simultaneously. Figure~\ref{fig:perfect-region} shows the values of $\epsilon$ in the complex plane for which the zeros intersect the locally perfect contour. In the interior of this region, no globally perfect contour exists. The boundary of this region and its exterior correspond to values of $\epsilon$ with well-behaved normalizing flows and therefore globally perfect contours of integration.

The study of Boltzmann factors with zeros is motivated by the case of the finite-density fermion sign problem, where the practice of integrating fermions out yields precisely these sorts of singularities of the action. In~\cite{ref:nf} it was shown that for a simplified (mean-field) model of lattice fermions in any number of space-time dimensions, perfect contours do not exist at finite chemical potential. However, neither the discussion above nor that primitive example are sufficient to rule out the existence of perfect contours for realistic lattice fermions. Whether or not such contours exist remains an important open question.

%\section{Conclusion}\label{con}\todo

\acknowledgments
Y.Y.~is supported by the U.S.~Department of Energy under Contract No.~DE-FG02-93ER-40762. S.L.~is supported by the U.S. Department of Energy, Office of Science, Office of Nuclear Physics program under Award Number DE-SC-0017905.

\end{document}